# Charge Transfer Database for Bio-molecule Tight Binding Model Derived from Thousands of Proteins

Hongwei Wang, Fang Liu, Tiange Dong, Likai Du*, Dongju Zhang, Jun Gao

[1]Hubei Key Laboratory of Agricultural Bioinformatics, College of Informatics, Huazhong Agricultural University, Wuhan, 430070, P. R. China

[2]Institute of Theoretical Chemistry, Shandong University, Jinan, 250100, P. R. China

*To whom correspondence should be addressed.

Likai Du: dulikai@mail.hzau.edu.cn

**Abstract**

The anisotropic feature of charge transfer reactions in realistic proteins cannot be ignored, due to the highly complex chemical structure of bio-molecules. In this work, we have performed the first large-scale quantitative assessment of charge transfer preference in protein complexes by calculating the charge transfer couplings in all 20×20 possible amino acid side chain combinations, which are extracted from available high-quality structures of thousands of protein complexes. The charge transfer database quantitatively shows distinct features of charge transfer couplings among millions of amino acid side-chains combinations. The knowledge graph of charge transfer couplings reveals that only one average or representative structure cannot be regarded as the typical charge transfer preference in realistic proteins. This data driven model provides us an alternative route to comprehensively understand the pairwise charge transfer coupling parameters based structural similarity, without any require of the knowledge of chemical intuition about the chemical interactions.

# 1. Introduction

Charge transfer is one of the simplest but fundamental reactions in life science. [1-7] The electron or hole transfer reactions are possible between donors and acceptors separated by a long distance, i.e. across protein-protein complexes. [6, 8-13] In biological molecules, the superexchange (tunneling) and hopping mechanism are commonly used to interpret charge transfer processes. [5-6, 14-16] The tunneling mechanism is a one step process which exhibits a strong distance dependence, while the hopping mechanism provides an explanation for electron or hole transfer across long distances. Although the driving force of charge transfer reactions are "encoded" in the thousands of known protein structures, "decoding" them is challenging because of the complexity of natural proteins. [17-20]

The building blocks of proteins are only the twenty L-amino acids, which are distinguished by their distinct side-chain structures. Bioinformatics scientists have payed much attention to depict the structural significance of these protein complexes, and a large number of biological databases were constructed to classify protein structures in the last decades. [21-28] And the growing amount of high-quality experimental (X-ray, NMR, cryo-EM) proteins structures have opened space to improve our theoretical understanding of biological charge transfer reactions. The relative abundance of various modes of amino acid contacts (van der Waals contacts, hydrogen bonds) could be completely exploited to understand the nature of electron transfer in proteins. Therefore, it becomes increasingly important to incorporate the available structural knowledge into our physical model development. [29-33]

Generally, in the bio-molecules charge transfer reactions, the charge transfer rate is proportional to the square of the donor/acceptor electronic coupling strength and the nuclear factor associated with the motion along the reaction coordinate. [6, 12-13, 34-35] Electronic coupling elements as an important component for biological charge transfer can be derived from various empirical or semi-empirical models [17, 35-39] and from direct electronic structure calculations. [40-45] For instance, Beratan *et. al.* introduces the graph theory to calculate the electronic coupling terms and search tunneling pathways or pathway families in bio-molecules, for which the electronic coupling is empirical and written as a product of a hypothetical closest contact terms, involving covalent, hydrogen bond, and van der Waals interactions. [35, 38, 46] Nowadays, the computations with more advanced models are becoming increasingly possible to obtain the charge transfer couplings for ensembles of structures. Therefore, it is desirable to go beyond the empirical models and directly calculate charge transfer coupling parameters for millions of molecular fragments.

In this work, we present a promising computational protocol to construct the charge transfer coupling database, which provides an overall view of electron transfer couplings among millions of amino acid side chain combinations. This database reveals that each type of amino acid combinations contains specific geometric distributions, and thus distinct charge transfer couplings populations. The possible structural changes could significantly influence the electrical properties in bimolecular fragments. The amino acid charge transfer database is large enough to sufficiently represent possible occurrences of amino acid contacts in realistic proteins. Thus, the pairwise charge transfer interactions among discretized libraries of amino acid side-chain conformations, as a powerful look-up table, enables us to directly obtain the overall charge transfer preferences for any structures, in the foreseen big data scenario.

## 2. Methods and Computational Details

### 2.1 Systematic Preparation for Amino Acid Side Chains Interaction Database

The Protein Data Bank (PDB) contains a wealth of data on nonbonded biomolecular interactions. This information is useful for us to develop a data driven or informatics based model. Here, the data collection procedure is accomplished by extracting the structural data from an improved version of the "Atlas of Protein Side-Chain Interactions", which are derived from thousands of unique structures of protein complexes[47]. As of June 2017, the Atlas comprised 482555 possible amino acid side-chain combinations for 20×20 sets of amino acid contacts. Each type of amino acid combinations has been carefully classified up to six clusters based on geometric similarity. In this fashion, each cluster has a clear-cut representitive structure. The procedure to extract each dimer complex has been described in the work of Singh and Thornton.[48]

Scheme 1 provides the workflow to derive the charge transfer couplings. The initial structures in the amino acid side-chain database contains only the coordinates of heavy atoms, and the missing hydrogens were added using the *tleap* module in AmberTools package[49]. The point of cutting covalent bond is saturated with hydrogen atoms (i.e., either the $C_\alpha$ or $C_\beta$ atom). In order to eliminate any potential nonspecific interactions, the positions of hydrogen atoms were optimized for each dimer at the semiempirical PM6 level with Gaussian 09 package[50]. The coordinates of the heavy atoms were kept fixed during the optimization procedure. The optimized structures are used for our subsequent construction of charge transfer couplings database and knowledge graph analysis.

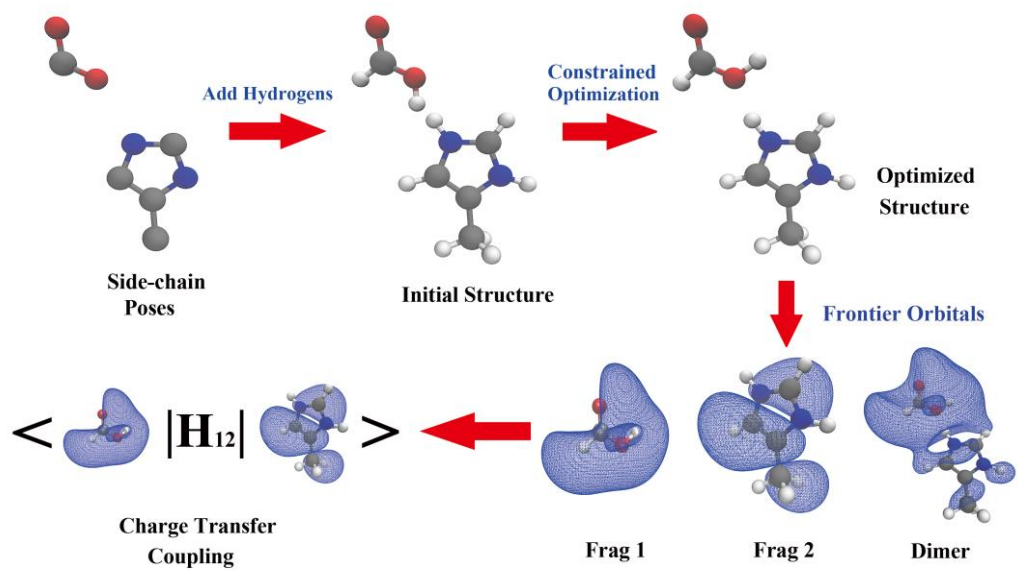

**Scheme 1**. The workflow of calculating the charge transfer coupling for any amino acid side-chain combination.

## 2.2 Charge Transfer Coupling Calculations

The charge transfer couplings for the amino acid side-chain combinations are derived from the ab initio calculations according to the concept of tight-binding approximation, following the previous work of Liu and co-workers[51-53]. The on-site energy and transfer integral are directly calculated as electronic coupling of the HOMOs (LOMOs) of the dimers. The derived expressions for the on-site energy and transfer integral can be written as

$$\varepsilon_n = \langle \phi_n | h | \phi_n \rangle = \langle \phi_n | -\frac{1}{2}\nabla^2 + \sum_L V_L | \phi_n \rangle \tag{1}$$

$$t_{n,n+1} = -\langle \phi_n | h | \phi_{n+1} \rangle = -\langle \phi_n | -\frac{1}{2}\nabla^2 + \sum_L V_L | \phi_{n+1} \rangle \tag{2}$$

In the above equations, $h$ is the electronic Hamiltonian. And $\phi_n$ and $\phi_{n+1}$ are the orbitals of

HOMOs (LUMOs) belonging to the amino acid fragment of the site $n$ and $n+1$, which can be obtained by solving the one-electron Schrödinger equation for one non-perturbation and isolated site.

$$\left[-\frac{1}{2}\nabla_i^2 + V_L\right]\phi_l = \varepsilon_l^0 \phi_l \tag{3}$$

whereas, $\phi_l$ is the orbital of the site L, and $\varepsilon_l^0$ is the corresponding orbital energy. Then, the molecular orbitals for the entire biomolecules can be expanded as the linear combination of site orbitals. The site potential $V_L$ can be given as,

$$V_L(i) = \sum_{a \in L} -\frac{Z_a}{r_{ai}} + \frac{1}{2} \sum_{j \in L, j \neq i} \frac{1}{r_{ij}} \tag{4}$$

The Eq. 3 can be solved by the self-consistent Hartree-Fock (HF) or DFT calculations. And the site potential matrix in Eq. 4 can be also obtained from the self-consistent field calculations. Both this matrix and HOMO (LUMO) of individual fragments were calculated at B3LYP/6-31G* level.

According to the tight-binding approximation, the electron belonging to site $n$ is mainly affected by the potential of site $n$ and its neighbors, and other sites show minor contributions. For the calculation of transfer integral, we only require the two-body site Fock matrix. Thus, the transfer integral can be written as,

$$t_{n,n+1} \approx -\langle\phi_n| -\frac{1}{2}\nabla^2 + V_n + V_{n+1} |\phi_{n+1}\rangle \tag{5}$$

Because the monomer orbitals of each site derived from self-consistent field calculations are non-orthogonal. An orthogonal basis set that maintains as much as possible the initial local character of the monomer orbitals can be obtained from Lowdin's symmetric transformation[51, 54-55]. The effective transfer integral can be written as,

$$t_{ij}^{eff} = \frac{t_{ij} - \frac{1}{2}(\varepsilon_i + \varepsilon_j)S_{ij}}{1 - S_{ij}^2} \tag{6}$$

In Eq 6, $S_{ij}$ is the overlap integral between orbital of the site i and j. And the transfer integral describes the ability to perform the charge transfer among neighbor sites.

**Results and Discussions**

First, the charge transfer integrals are calculated for millions of amino acid side-chain combinations to reveal the richness of biological charge transfer reactions in realistic proteins. This database is helpful to describe how the conformation ensemble influences charge transfer couplings within distinct protein structures. Here, without losing any generality, we will restrict our study to electron hole transfer coupling between HOMOs of each amino acid side-chain combinations. To facilitate our following discussions, the words "hot" and "cold" are used to describe the residues with larger and smaller charge transfer coupling values.

Traditionally, the charge transfer coupling was often assumed to be a constant or an empirical formula for each type of residue combinations in modelling realistic bio-molecules systems.[35, 38, 46, 51-53] Thus, Table 1 presents the average charge transfer couplings as constants of the overall 20×20 possible amino acid pairs, which reveals that most charge transfer couplings are non-zero and below 0.05 eV. The twenty amino acids can be divided into several groups according to the chemical compositions of their side chains, i.e. non-polar/hydrophobic groups, non-polar/aromatic groups, polar/neutral groups, polar/basic groups and polar/acidic groups. In general, the transfer couplings for the polar/polar or basic/acidic combinations are greater than the pairwise hydrophobic combinations[56], with only minor exceptions.

**Table 1**. The average transfer couplings matrix for the 20×20 possible pairs over their representative structures. The filled colors are used to distinguish residue types. The white refers to non-polar, purple refers to aromatic, green refers polar and neutral, blue refers to polar and basic and red refers to polar and acidic amino acids. Note, the significant charge transfer couplings are highlighted with blue (0.05－0.1 eV) and red (≥0.10 eV) fonts.

|  | GLY | ALA | VAL | LEU | ILE | PRO | PHE | TRP | TYR | SER | CYS | MET | ASN | GLN | THR | LYS | ARG | HIS | ASP | GLU |
|---|---|---|---|---|---|---|---|---|---|---|---|---|---|---|---|---|---|---|---|---|
| GLY | 0.03 | 0.02 | 0.02 | 0.01 | 0.02 | 0.03 | 0.03 | 0.02 | 0.03 | 0.03 | 0.01 | 0.06 | 0.04 | 0.02 | 0.03 | 0.02 | 0.02 | 0.11 | 0.08 | 0.08 |
| ALA | 0.03 | 0.02 | 0.02 | 0.02 | 0.02 | 0.02 | 0.03 | 0.01 | 0.02 | 0.03 | 0.02 | 0.06 | 0.04 | 0.01 | 0.03 | 0.03 | 0.04 | 0.05 | 0.05 | 0.02 |
| VAL | 0.02 | 0.02 | 0.07 | 0.03 | 0.03 | 0.02 | 0.01 | 0.02 | 0.05 | 0.02 | 0.05 | 0.02 | 0.04 | 0.02 | 0.02 | 0.08 | 0.05 | 0.08 | 0.04 | 0.01 |
| LEU | 0.01 | 0.02 | 0.07 | 0.04 | 0.03 | 0.02 | 0.05 | 0.04 | 0.03 | 0.03 | 0.03 | 0.08 | 0.01 | 0.02 | 0.01 | 0.03 | 0.03 | 0.08 | 0.03 | 0.02 |
| ILE | 0.02 | 0.02 | 0.04 | 0.01 | 0.05 | 0.01 | 0.02 | 0.02 | 0.06 | 0.03 | 0.03 | 0.04 | 0.03 | 0.02 | 0.02 | 0.07 | 0.04 | 0.04 | 0.02 | 0.02 |
| PRO | 0.02 | 0.01 | 0.01 | 0.03 | 0.01 | 0.01 | 0.02 | 0.05 | 0.03 | 0.01 | 0.03 | 0.04 | 0.01 | 0.01 | 0.03 | 0.02 | 0.03 | 0.04 | 0.02 | 0.01 |
| PHE | 0.02 | 0.02 | 0.04 | 0.03 | 0.02 | 0.03 | 0.02 | 0.01 | 0.04 | 0.02 | 0.02 | 0.01 | 0.01 | 0.01 | 0.01 | 0.03 | 0.16 | 0.05 | 0.02 | 0.01 | 0.01 |
| TRP | 0.04 | 0.01 | 0.02 | 0.04 | 0.03 | 0.05 | 0.01 | 0.02 | 0.01 | 0.05 | 0.04 | 0.00 | 0.03 | 0.01 | 0.02 | 0.09 | 0.06 | 0.06 | 0.01 | 0.02 |
| TYR | 0.01 | 0.02 | 0.03 | 0.03 | 0.01 | 0.04 | 0.02 | 0.01 | 0.02 | 0.02 | 0.03 | 0.01 | 0.01 | 0.01 | 0.01 | 0.08 | 0.03 | 0.02 | 0.02 | 0.03 |
| SER | 0.06 | 0.05 | 0.04 | 0.06 | 0.02 | 0.06 | 0.03 | 0.04 | 0.02 | 0.08 | 0.00 | 0.00 | 0.06 | 0.04 | 0.05 | 0.05 | 0.02 | 0.01 | 0.11 | 0.11 |
| CYS | 0.00 | 0.03 | 0.05 | 0.11 | 0.04 | 0.08 | 0.02 | 0.02 | 0.03 | 0.06 | 0.34 | 0.00 | 0.00 | 0.04 | 0.04 | 0.00 | 0.04 | 0.03 | 0.00 | 0.00 |
| MET | 0.02 | 0.03 | 0.02 | 0.04 | 0.03 | 0.05 | 0.06 | 0.02 | 0.04 | 0.07 | 0.01 | 0.03 | 0.00 | 0.02 | 0.01 | 0.06 | 0.03 | 0.05 | 0.06 | 0.03 |
| ASN | 0.04 | 0.03 | 0.03 | 0.04 | 0.02 | 0.01 | 0.03 | 0.02 | 0.03 | 0.04 | 0.08 | 0.00 | 0.06 | 0.04 | 0.08 | 0.03 | 0.01 | 0.05 | 0.06 |
| GLN | 0.02 | 0.03 | 0.02 | 0.01 | 0.00 | 0.01 | 0.02 | 0.03 | 0.02 | 0.14 | 0.00 | 0.00 | 0.09 | 0.10 | 0.06 | 0.09 | 0.04 | 0.04 | 0.07 | 0.10 |
| THR | 0.05 | 0.04 | 0.03 | 0.02 | 0.01 | 0.04 | 0.03 | 0.01 | 0.02 | 0.07 | 0.02 | 0.01 | 0.07 | 0.04 | 0.10 | 0.04 | 0.04 | 0.02 | 0.13 | 0.12 |
| LYS | 0.01 | 0.01 | 0.06 | 0.03 | 0.03 | 0.19 | 0.11 | 0.07 | 0.05 | 0.02 | 0.00 | 0.08 | 0.01 | 0.03 | 0.04 | 0.03 | 0.06 | 0.06 | 0.03 | 0.02 |
| ARG | 0.01 | 0.04 | 0.05 | 0.01 | 0.05 | 0.00 | 0.00 | 0.03 | 0.06 | 0.00 | 0.00 | 0.00 | 0.02 | 0.04 | 0.02 | 0.09 | 0.07 | 0.08 | 0.05 | 0.04 |
| HIS | 0.03 | 0.03 | 0.05 | 0.04 | 0.03 | 0.01 | 0.11 | 0.06 | 0.05 | 0.01 | 0.04 | 0.09 | 0.02 | 0.01 | 0.03 | 0.00 | 0.05 | 0.03 | 0.01 | 0.01 |
| ASP | 0.10 | 0.06 | 0.02 | 0.01 | 0.03 | 0.02 | 0.00 | 0.03 | 0.02 | 0.10 | 0.00 | 0.00 | 0.00 | 0.05 | 0.07 | 0.07 | 0.02 | 0.02 | 0.01 | 0.01 |
| GLU | 0.04 | 0.09 | 0.02 | 0.02 | 0.01 | 0.02 | 0.01 | 0.04 | 0.03 | 0.19 | 0.00 | 0.00 | 0.05 | 0.10 | 0.19 | 0.02 | 0.06 | 0.01 | 0.01 | 0.14 |

The knowledge graph is applied to visualize charge transfer couplings for 20×20 possible amino acid combinations. In Figure 1a, the edge weights in our graph are assigned as the average values of the unsigned charge transfer couplings between two types of aminio acids (see Table 1). Each vertex represents an amino acid. The edge width between the nodes is linearly corresponding to the absolute value of the charge transfer couplings. This graph representation provides a self explanatory of the significant charge transfer interactions among amino acids, such as the Glu/Ser, Lys/Pro and *so on*. Note that, the remarkable charge transfer couplings between cysteine residues are mainly caused by the possible disulfide bonds in realistic proteins.

In Figure 1b, we try to go beyond the analysis of average charge transfer couplings, and details of the entire 482555 amino acids pairs were exploited. The edge weights are assigned as the number of structures with charge transfer couplings larger than 0.05 eV in each kind of amino acid combinations. This graph retains most topological features of charge transfer couplings in Figure 1a, which indicates that the average structures of each amino acid pair provide a reasonable approximation to interpret the charge transfer couplings in realistic proteins. Figure 1c summarizes the charge transfer significance of each amino acid, which is consistent with our common sense that most remarkable residues for charge transfer reactions are those polar residues, such as Ser and Lys. This one dimensional representation may be useful as a reference material to qualitatively understand the possible charge transfer features in proteins.

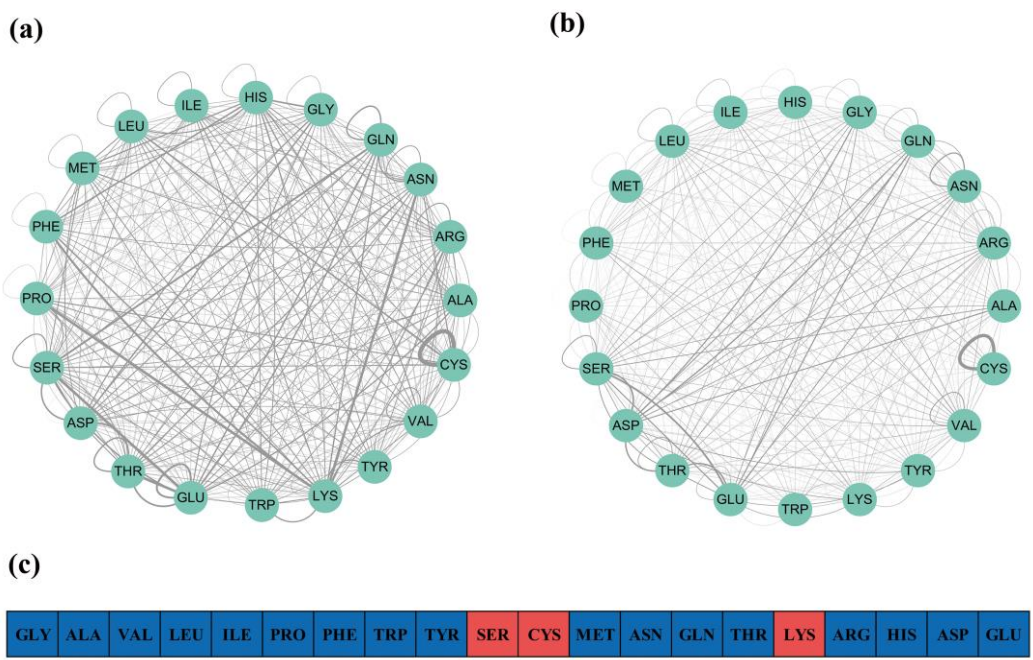

**Figure 1.** Knowledge graph of the charge transfer couplings among residues. (a) The charge transfer knowledge graph connected by their average charge transfer couplings. (b) The number distribution graph connected by the number of amino acid pairs with charge transfer couplings larger than 0.05 eV. (c) The residues with significant charge transfer couplings are highlighted with blue (<0.05 eV) and red (≥0.05 eV) backgrounds.

Next, we focus on the top 16 amino acid pairs with significant charge transfer couplings in 400 possible amino acid pairs. In Figure 2, the box plots are used as a quick way of examining the variation in statistical population of charge transfer couplings. Each box plot refers to the distinct geometric clusters for a specific type of the amino acid pair. The median value of the charge transfer couplings for each geometric cluster is significantly different. The spacings between the different parts of the box indicate the degree of dispersion (spread) and skewness of the data. The overall charge transfer couplings are widely distributed among geometric clusters within a specific amino acid combination. In summary, each type of amino acid pair or even each geometric cluster may contain widespread charge transfer couplings. Thus, it is not always a good choice to assume the charge transfer coupling is the same for one type of amino acid combinations in realistic proteins and their dynamics studies.

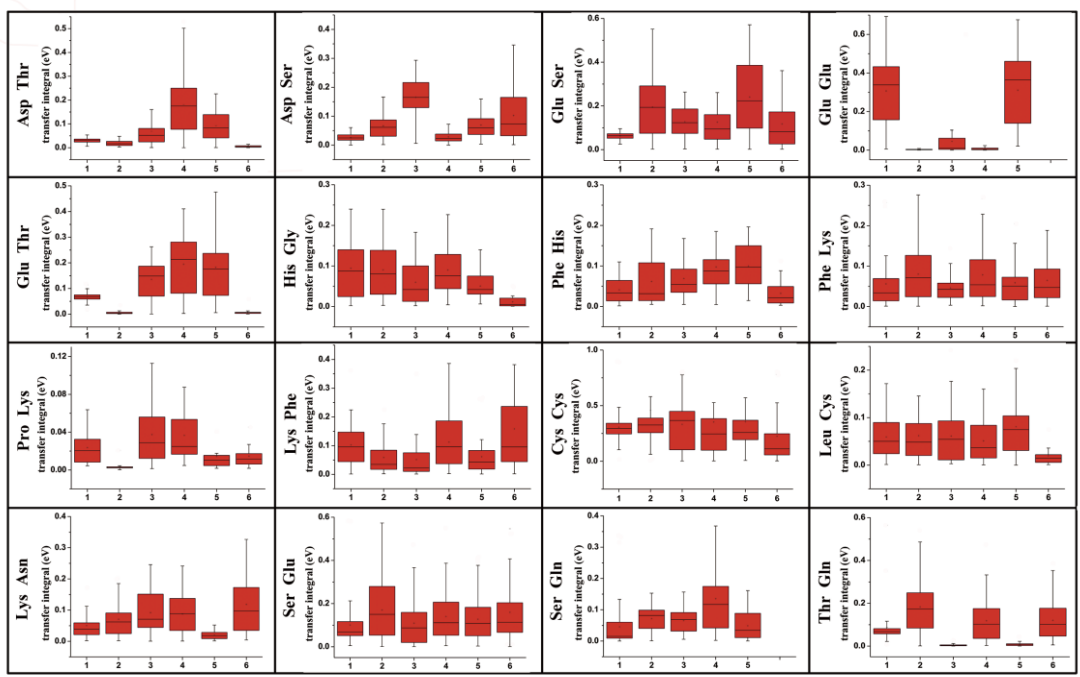

**Figure 2**. Box plots are used to depict the character of charge transfer couplings. Each plot shows the distinct clusters for a specific amino acid pair. The bottom and top of the box are the first and third quartiles, and the band inside the box is the second quartile (the median). Note that, the scale of y-axis is not the same to enhance the comparison within each box plot.

Figure 3 provides three dimensional structure distributions for the Glu/Glu and His/Glu pairs as typical systems, which have an intermediate total number of observed contacts (2475 and 2583) in the protein amino acid side-chain altlas. Figure 3a and 3b show the geometric distributions for each amino acid pair, for which the amino acids have distinct interaction patterns, indicating their packing is not entirely random. [57-59] The population of charge transfer couplings are "encoded" in various model of geometric contacts, i.e. the hydrogen bonds or van der Waals contacts. The geometric distributions with charge transfer couplings greater than 0.05 eV are shown in Figure 3c and 3d. A few geometric clusters completely disappear in the Glu/Glu and His/Glu pairs, which indicates their minor contribution to the possible charge transfer events. These results suggest that the charge transfer couplings distribution of overall geometric clusters cannot be simply described by only one representative structure. And one must pay close attention when dealing with the geometric ensemble of amino acid pairs in realistic proteins, and the appropriate transfer coupling parameters should be applied only after performing tests on similar geometric features.

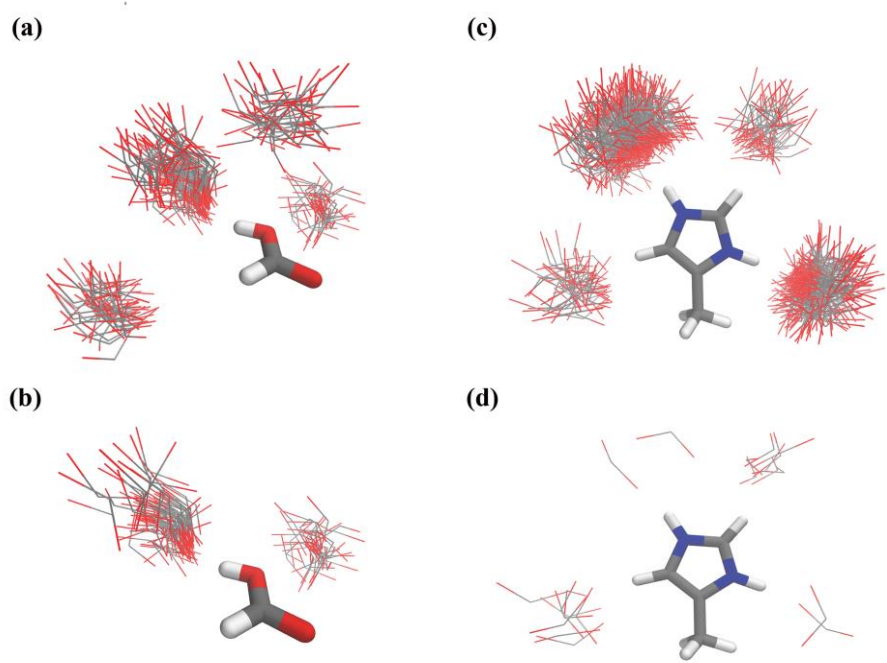

**Figure 3.** Example distribution and its associated geometric clusters for Glu/Glu (a) and His/Glu (b) pairs. The selective structural distributions for Glu/Glu (c) and His/Glu (d) pairs with the charge transfer coupling greater than 0.05 eV.

As mentioned above, the charge transfer couplings parameters are found to be very sensitive to the structural orientation of the amino acid pairs[3, 60-62], in the context of overall geometric distribution. The relative abundance of various modes of amino acid contacts leads to very different charge transfer couplings distribution. Thus, several geometric clusters can be assigned as "hot" contacts, while the others can be assigned as "cold" contacts. The plots of charge transfer couplings distribution of all amino acid pairs are available at http://github.com/dulikai/bidiu.

In Figure 4, the charge transfer couplings population is summarized for the Ser/Glu pair, which can be classified into six geometric clusters. The Ser/Glu pair is used as an illustrated example, with enough geometric contacts (3277 contacts) and charge transfer couplings strengths. First, the selected cluster representatives (red lines) are not always relate to the largest peak position in the charge transfer coupling distributions. And the cluster representative structure may even represent the extreme charge transfer coupling value in a few cases. Second, the charge transfer couplings distribution curves usually exhibit more than one peak, beyond the peak position near the zero value. Third, the charge transfer couplings distribution can be widespread in the same geometric clusters, although the RMSD of these structures in each cluster are only within 1.5 Å[48]. In summary, further geometric variables should be applied to measure the the charge transfer couplings distribution in realistic proteins.

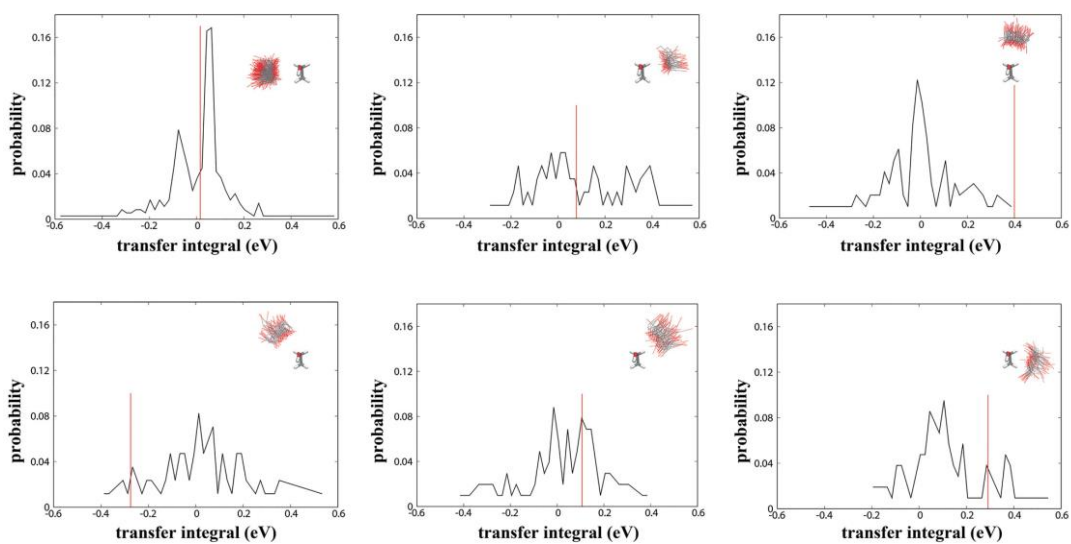

**Figure 4.** The distribution of charge transfer couplings (eV) for the six geometric clusters of Ser/Glu pair are shown, whereas Ser is the center fragment. The red line refers to the charge transfer couplings for the representative structure of each geometric cluster.

The explicit geometric correlation with the charge transfer couplings is investigated. The electronic coupling is usually supposed to decay exponentially with the distance between donors and acceptors [63-65] Figure 5 shows the charge transfer couplings along the center to center distance between the two amino acid pairs. Due to the anistropic feature of amino acids, it is hard to say that the amino acid pairs with larger charge transfer couplings show relatively shorter pairwise distances. In most case, the values of charge transfer couplings can differ by one order of magnitude at the median range (4.0~5.0 A). The charge transfer couplings remarkably decay to nearly zero beyond the median range. In addition, the overall charge transfer couplings distribution is strongly related to the physicochemical properties of amino acids. And each type of amino acid pairs exhibits its specific charge transfer couplings feature or fingerprint.

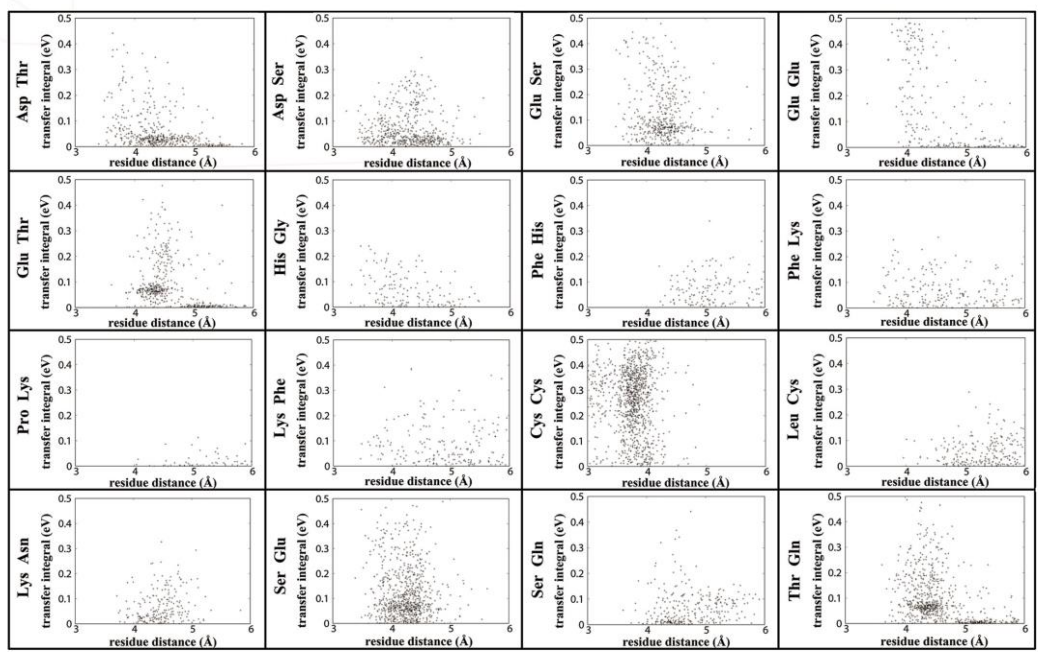

**Figure 5**. The charge transfer couplings as a function of amino acid pairwise distance for top 16 amino acid pairs.

It is expected more abundant electronic information can be derived from the orientation of amino acid side-chain contacts. Figure 6 provides the charge transfer couplings distribution as a function of distance and angle between pairwise amino acids. The overall view of heat maps suggests that each type of amino acid combinations contains specific its charge transfer couplings distribution. And the charge transfer coupling shows strong anisotropic features among various clusters of amino acid side-chains combinations. We can find a few distinct "hot" regions, for which the unsigned charge transfer couplings are larger than its surroundings. Thus, the angular variable should be considered in the analytical estimate of the charge transfer couplings among amino acids. The anistropic stacking of protein side chains could be a critical factor to determine the electronic properties of proteins. And we suggest that the possible structural changes could significantly influence the electronic properties in proteins. [60, 62, 66-67]

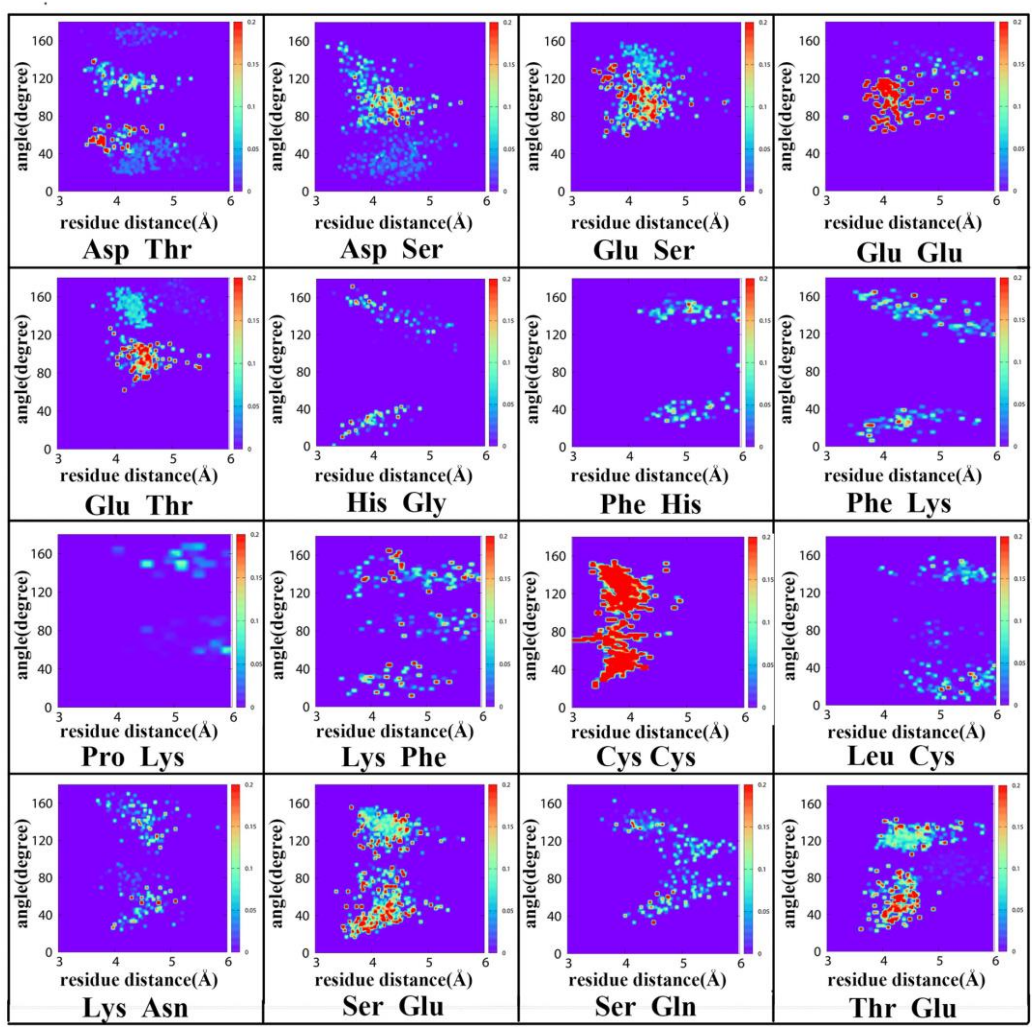

**Figure 6**. The distance *v.s.* angular dependent charge transfer couplings for top 16 amino acid pairs.

Finally, we present the selected structures for top 16 amino acid pairs with most significant charge transfer couplings in Figure 7. It is clear that the amino acid pair should have specific contacts to achieve high charge transfer couplings. Various inter-molecular interactions can be responsible for the significant charge transfer reactions. The hydrogen bonds are most common in the available amino acids pairs, such as Asp/Thr, Asp/Ser, Glu/Ser, Glu/Glu, Glu/Thr, and *so on*. This is consistent with our common sense[3, 20, 46, 56]. The π⋯π stacking or C-H⋯π interactions between the hydrophobic amino acid side-chain and aromatic rings are also observed to be important for the charge transfer reactions (i.e. Lys/Phe, Phe/His, and *so on*), although their absolute couplings values are not very large (~0.05 eV). The role of C-H⋯π interactions in charge transfer reactions is not commonly recognized[68-69], although the C-H···π interactions are reported to play an important role due to their significant occurances in organic crystals, proteins and nucleic acids.[70-72] Further investigation of the role of C-H···π interactions in charge transfer interactions is needed. In summary, the exploitation of the mega data sets allows us to rationalize the charge transfer couplings and its structural characters on the same foot.

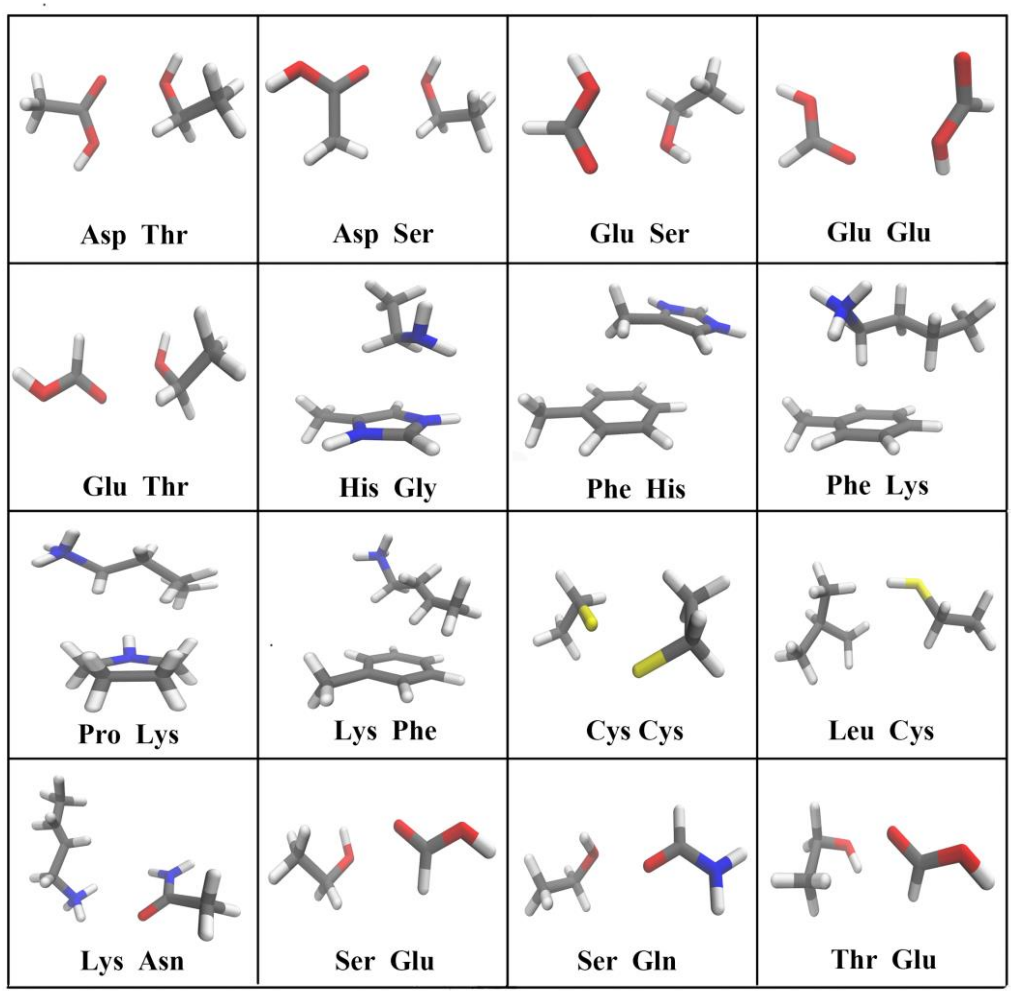

**Figure 7.** The selected structures for the top 16 amino acid pairs with the most significant charge transfer couplings.

**Conclusions**

In summary, we present a promising protocol to construct a charge transfer database at residue level, which are derived from millions of electronic structure calculations among 20×20 possible amino acid side-chains combinations. In this fashion, the possible charge transfer properties among residues could be understood in an more explicit and more intuitionistic fashion, without any require of the knowledge of chemical intuition about the chemical interactions or empirical formulas. And the possible structural changes could significantly influence the electronic properties in proteins. Based on these observations, we suggest that the protein charge transfer can be accomplished by the selective arrangement of interacting amino acids orientations.

The construction of charge transfer database for amino acids presents one of the key steps towards understanding the electronic structure information in proteins. Future work may be possible to enumerate the most common "hot" motif that are suitable for charge transfer reactions in proteins by reusing sophisticated charge transfer parameters.

## Acknowledgements

The work is supported by National Natural Science Foundation of China (Nos. 21503249, 21373124), and Huazhong Agricultural University Scientific & Technological Self-innovation Foundation (Program No.2015RC008), and Project 2662016QD011 and 2662015PY113 Supported by the Fundamental Founds for the Central Universities. The authors also thank the support of Special Program for Applied Research on Super Computation of the NSFC-Guangdong Joint Fund (the second phase) under Grant No.U1501501.